\documentclass[final,3p,times]{elsarticle}

\usepackage{epsfig}

\usepackage{amssymb}

\biboptions{compress}

\usepackage{amsmath}

\usepackage{latexsym}

\usepackage{amsbsy}

    \usepackage{subfig}
    \usepackage{booktabs}
    \usepackage{epsfig}
    \usepackage{hyphenat}
    \usepackage{multicol}

\journal{Physica A}

\begin{document}

\begin{frontmatter}



\title{Quantum Brownian motion model for the stock market}


\author[label1,label3]{Xiangyi Meng}
\author[label2]{Jian-Wei Zhang}
\author[label1]{Hong Guo\corref{CorrespondingAuthor}}
\cortext[CorrespondingAuthor]{Corresponding author. Tel.: +86 10 6275 7035}
\address[label1]{State Key Laboratory of Advanced Optical Communication$\text{,}$ Systems and Networks$\text{,}$ School of Electronics Engineering and Computer Science$\text{,}$ and Center for Quantum Information Technology, Peking University, Beijing 100871, China}
\address[label2]{School of Physics, Peking University, Beijing 100871, China}
\address[label3]{Department of Physics, Boston University, Boston, Massachusetts 02215, USA}
\ead{hongguo@pku.edu.cn}
\date{}
\begin{abstract}
It is believed by the majority today that the efficient market hypothesis is imperfect because of market irrationality. Using the physical concepts and mathematical structures of quantum mechanics, we construct an econophysics framework for the stock market, based on which we analogously map massive numbers of single stocks into a reservoir consisting of many quantum harmonic oscillators and their stock index into a typical quantum open system--a quantum Brownian particle. In particular, the irrationality of stock transactions is quantitatively considered as the Planck constant within Heisenberg's uncertainty relationship of quantum mechanics in an analogous manner. We analyze real stock data of Shanghai Stock Exchange of China and investigate fat-tail phenomena and non-Markovian behaviors of the stock index with the assistance of the quantum Brownian motion model, thereby interpreting and studying the limitations of the classical Brownian motion model for the efficient market hypothesis from a new perspective of quantum open system dynamics.
\end{abstract}


\end{frontmatter}


\section{Introduction}
\label{Section1}
Since varied concepts and methods in statistical physics have been successfully introduced to economics with especial thanks to the pioneering work of Mantegna and Stanley in the 1990s~\cite{Stanley2,Stanley}, econophysics has gained momentum ever since and generated countless powerful tools to merge statistical physics and economics~\cite{EconophysicsIsNotIrrelevance}. The early-stage elementary research has hitherto experienced widely development and conducted multidirectional studies in, for example, scaling and power-law behaviors of various markets~\cite{ScalingBehavior1, ScalingBehavior2, paper1b, ZJW3}, fractality/multifractality~\cite{Fractal, FractalMarkets, Multifractal1, PriceLimit1} in financial indices, complex networks~\cite{ComplexNetworks, ComplexNetworks1, PerturbationsInComplexNetworks} for dynamics of microeconomics, etc. Even from a general physical perspective, natural concepts and elaborate microscopic models can both help to explore fundamental economic issues of market efficiency and stability~\cite{PowerMarketAmplifyFluctuation, SimpleMarketModelsFail, EfficiencyAndStability, DescriptionOfEMH, PrecariousnessAndEconomicStratification}. Recently, integration of quantum mechanics into finance has been seriously attempted. No sooner had path integral methods first been applied to options pricing successfully~\cite{QuantumFinancePathIntegrals} than a series of financial analyses with concepts and tools from quantum mechanics used were launched~\cite{PersistentFluctuationsStockMarkets, PRE1, QuantumStatisticalStockMarkets, FuDanQuantumModel, MinimalLengthUncertainty, QF4, QF2, QF1, QSPHM}. Despite of the mathematical success of quantum-mechanics models for financial instruments, few studies have been tried to exploit quantum statistical dynamics relying on open-system concepts yet. Among standard financial instruments, the stock market is the most symbolic and is critically concerned at all time. The study of the stochastic dynamics of stock markets was first systematically directed by Fama~\cite{Fama1}. In his study, use of the Brownian motion (random walk) was made to explain the unpredicability of stock market prices, which was polished to establish the remarkable efficient market hypothesis (EMH)~\cite{Fama1}. The rationale behind the principal idea of complete market rationality in the EMH, however, is suspected by the majority today. As negative empirical evidence such as non-Markovian memory~\cite{paper1c, ZJW1} and fat-tail deviation~\cite{NonGaussianPDF, ZJW4} is commonly found, it is obvious that the stock market does not satisfy the classical Brownian motion model (cBm(m)). It is thus believed by behavioral economists that the stock market is certainly affected by market irrationality. Existed econophysical models are more often involved with sophisticated quantities. Yet the quantification of market irrationality remains an interesting topic, of which the consideration should be more conceptual than over-mathematical. The criticism of applying multifractal analyses to intraday stock market indices~\cite{MultifractalModelIsWrong}, as an example, implies a lapse of concentration in overusing mathematical tools other than identifying instructive physical concepts for economics.

In 1933, Frisch presented a damped harmonic oscillator model~\cite{Frisch}. As one of the most influential models in dynamic economics, it presumes that on account of stock trading and its auxiliary expense, the price of a single stock should oscillate and dissipate to equilibrium like a damped harmonic oscillator while being impelled by time-varying information outside~\cite{QSPHM}. Notwithstanding its intuitive physical point of view, Frisch's model cannot explain why there always exists persistent fluctuations of the stock price~\cite{PingChen1}. After an initial impulse, the oscillation intensity of a classical damped harmonic oscillator should rapidly decay exponentially and reach equilibrium~\cite{DampedOscillationIsNotPersistent}. The equilibrium price should reflect the real value of the stock. However, empirical data reveal that the volatility of a single stock exists even if there is no information to impel the stock price~\cite{PersistentFluctuationsStockMarkets}. It is suggested in Ref.~\cite{PersistentFluctuationsStockMarkets} that one can use a quantum damped harmonic oscillator model instead of a classical one to describe price fluctuations of a single stock, with the price probability distribution described by a quantum wave function. For the stock price volatility $\sim \hbar$, it is natural for us to re-identify the concept of the Planck constant in dynamic economics. As we will find later, $\hbar$, as a source of additional uncertainty, can possibly be identified as the irrationality of stock transactions in the stock market. It may thus play a pivotal role of bridge between dynamic economics and behavioral economics.

The rest of the paper is organized as follows. In Section~\ref{BrownianMotionInTheStockMarket}, the EMH with its limitations is examined using the data of Shanghai Stock Exchange of China. A quantum Brownian motion model (qBm(m)) and its mapping to the stock market are then introduced in details in Section~\ref{QuantumBrownianMotionModel}. Moments and non-Markovian autocorrelations of the qBm(m) for the stock market regarding fat-tail non-Gaussian distribution and non-Markovian memory are later calculated and analyzed in Sections~\ref{MomentsOfTheQuantumBrownianMotionModel}~and~\ref{NonMarkovianFeaturesOfTheQuantumBrownianMotionModel}. Discussion and conclusion are finally given in Section~\ref{Conclusion}.

\section{Brownian motion in the stock market}
\label{BrownianMotionInTheStockMarket}
To begin with, we introduce the cBm(m) for the stock market with its mathematical description. Under the mathematical description of the classical Brownian motion, a one-dimensional free Brownian particle with its stochastic dynamics raised by the fluctuation and dissipation of environment satisfies a random walk process after a long time $t$. Its coordinate $x(t)$, as a random variable, has a Gaussian probability distribution with variance ${\sigma }^{2} t$, where ${\sigma }^{2}$ equals $kT/M\gamma$, with respect to the mass of the particle $M$, dissipation coefficient $\gamma$, and temperature $kT$ of the environment. The probability distribution function of the Brownian particle in the phase space is generally described by the Markovian Klein-Kramers equation~\cite{KKEquation}
\begin{equation}\label{KKeqn}
\frac{\partial }{\partial t}\mathcal{P}(x,p,t)=-\frac{p}{M}\frac{\partial }{\partial x}\mathcal{P}+2\gamma \frac{\partial }{\partial p}\left( p\mathcal{P} \right)+2\gamma MkT\frac{{{\partial }^{2}}}{\partial {{p}^{2}}}\mathcal{P},
\end{equation}\\
in which the second-order partial differential terms contribute to a Gaussian-shape solution~\cite{DifferentialEquations}. Given the momentum $p(t)=M dx/dt$, we know that $p(t)/M$ is a white noise process and its autocorrelation function $R(\tau )$ is $\left\langle p\left( t+\tau  \right)p\left( \tau  \right) \right\rangle /{{M}^{2}}={{\sigma }^{2}}\delta (\tau )$ with respect to time interval $\tau$, proving that the cBm(m) is a Markovian process. For economic studies, the classic concept of EMH presumes that the motion of a stock market index $S(t)$ is a stochastic process, i.e., $d(\ln S)=\left(\mu -{\sigma }^{2}/2 \right) dt+\sigma dW$. $\mu$ and $\sigma$ represent the drift rate and volatility of the stock index. $\sigma dW$ is a classical random-walk process.

\begin{figure}[thbp]
\begin{centering}
\subfloat[ ]{\begin{centering}
\includegraphics[height=4.8cm]{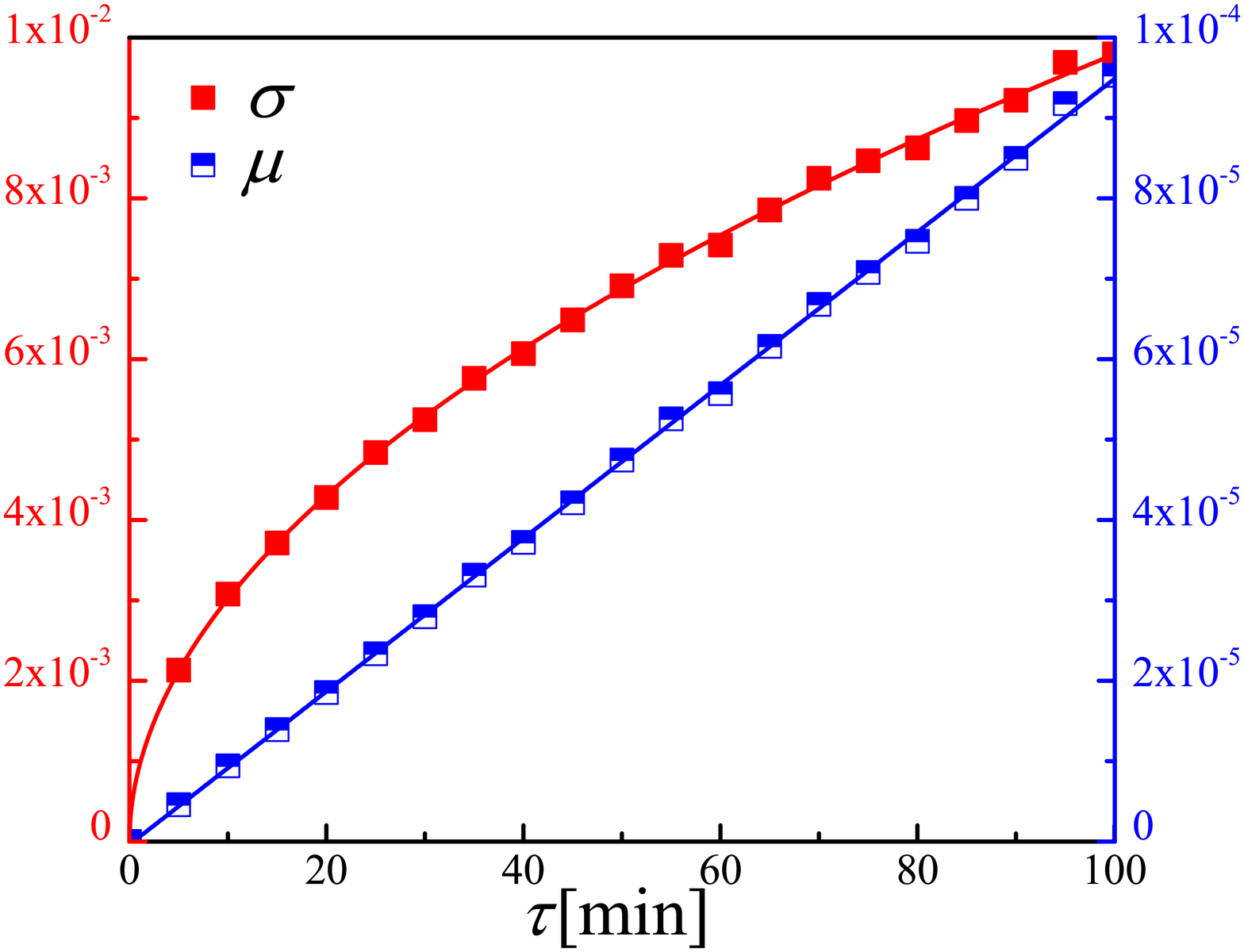}
\par\end{centering}
}\subfloat[ ]{\begin{centering}
\includegraphics[height=4.8cm]{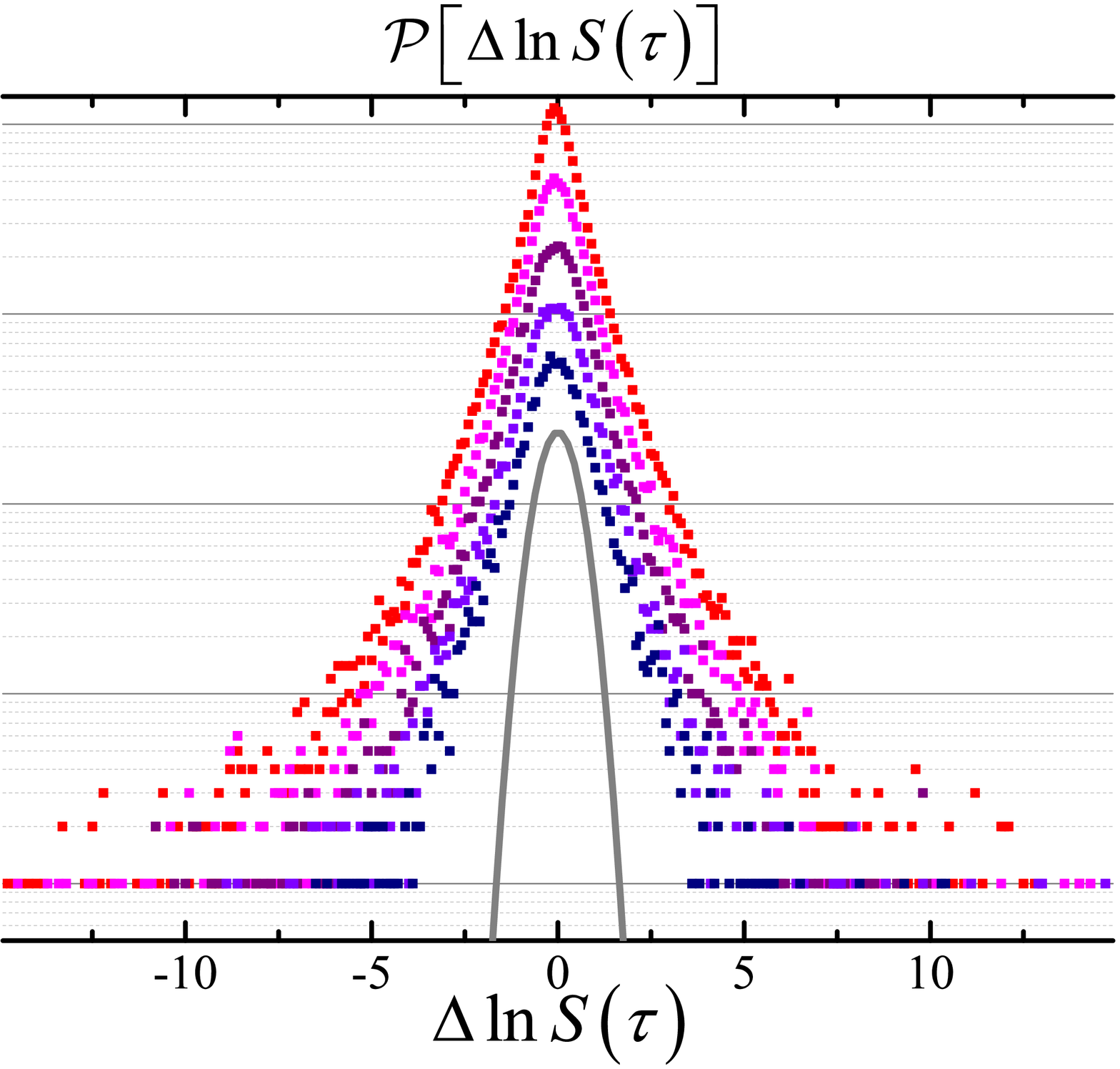}
\par\end{centering}
}\subfloat[ ]{\begin{centering}
\includegraphics[height=4.8cm]{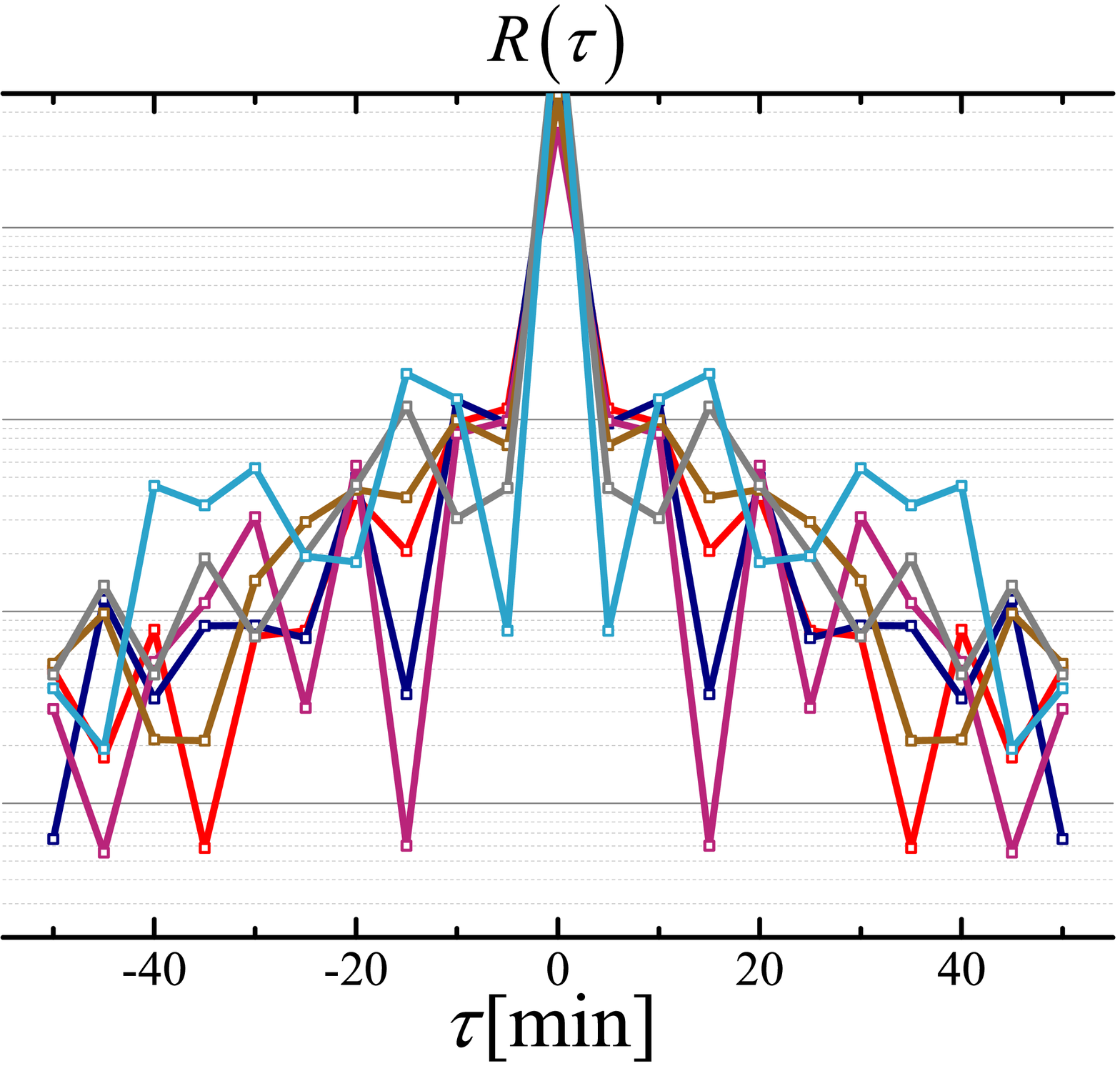}
\par\end{centering}
}
\par\end{centering}
\caption{\label{Fig1}Features of the Shanghai Composite Index of China from 1999 to 2013: (a) drift rate (blue line) and volatility (red line) derived from the $5,10,15,...,100\min$ lines of the index; (b) frequency distributions of the stock index return $\Delta \ln S(\tau)=\left[ \ln S(t+\tau)-\ln S(t) \right]/\tau$ (the drift term is already eliminated) with $\tau=5,10,20,40,80\min$ from top to bottom followed by the Gaussian distribution; (c) autocorrelation features of the stock index return with regard to six arbitrarily chosen time periods in 1999-2013. (For interpretation of the references to color in this figure legend, the reader is referred to the web version of this article.)}
\end{figure}

To test the validity of EMH for real stock indices, we analyze the $5,10,15,...,100\min$ lines data of the Shanghai Composite Index (SCI) of China from 1999 to 2013. It is obviously shown in Fig.~\ref{Fig1}(a) that $\sigma$ and $\mu$ are proportional to ${\tau}^{1/2}$ and $\tau$, respectively. Dynamics of the stock market index seems to obey the presumed stochastic process. However, in Fig.~\ref{Fig1}(b), the probability distribution of the stock index return (i.e., the difference of two successive data points of the logarithm of the stock index) does not fit the Gaussian distribution function but deviates with a fat tail, while this deviation disappears as $\tau$ increases. It is therefore implied that the presumption of Gaussian distribution is only valid in the long-term limit. In addition, Fig.~\ref{Fig1}(c) shows that the autocorrelation function $R(\tau)$ of the stock index is not a perfect Dirac-peak function at $\tau=0$ but is non-zero even when $|\tau| >20\min$. This time analysis indicates that the stock market is non-Markovian and thus convinces us again the cBm(m) is imperfect.

\section{Quantum Brownian motion model}
\label{QuantumBrownianMotionModel}
The description of the price of a single stock based on quantum mechanics provides an instructive point of view to deal with dynamical problems in the stock market~\cite{PersistentFluctuationsStockMarkets, QSPHM}. Under the quantum-mechanics-based description, one single stock $i$ is described by a wave function $\left| {{\psi }_{i}} \right\rangle$. The $X$ and $P$ representations are related to the (logarithmic) price and the trend of price, which are then considered in mathematics as the canonical coordinate and momentum of the stock, i.e., ${x}_{i}=\ln {{s}_{i}}$ and ${p}_{i}={{m}_{i}}d\left(\ln {{s}_{i}}\right)/dt$, respectively. The parameter ${m}_{i}$ is the inertia of the stock price against possible trend and may be treated typically as the capital of the stock~\cite{QSPHM}. The probability distribution $\left\langle {{\psi }_{i}} | x \rangle\langle x | {{\psi }_{i}} \right\rangle = |\psi _{i}(x,t)|^{2}$ describes the probability density of measuring the price with result $x$. Due to the uncertainty relation $[X,P] = i\hbar$ in quantum mechanics, the more we learn of the coordinate the less we know the momentum (and vice versa)~\cite{QuantumMechanicsCT}. This original idea is applied to stocks because the more we know the stock price the less information we can use to estimate the trend of it~\cite{PersistentFluctuationsStockMarkets}. In actual transaction, we can only acquire knowledge of the distribution of possible trading price in a certain range and thus cannot measure the real value of the stock exactly. Complementarily, we can to some extent estimate the trend of the price by analyzing its distribution~\cite{PersistentFluctuationsStockMarkets}. In a more detailed sense, the single stock price is treated as a quantum harmonic oscillator with its Hamiltonian ${{H}_{i}}={{P}^{2}}/2{{m}_{i}}+{{m}_{i}}{{\omega }_{i}}^{2}{{X}^{2}}/2$ a closed-system one, where ${\omega }_{i}$ is the characteristic oscillating frequency of the stock~\cite{PersistentFluctuationsStockMarkets, QSPHM}. The probability distribution of the ground state $|\psi _{0}(x,t)|^{2}$ is not a Dirac-peak distribution function. In other words, the stock price will remain a small uncertainty of fluctuations even if there is no exterior influence on the stock. This revealed phenomenon cannot be explained by the classical oscillation theory for stocks~\cite{Frisch} but can be interpreted by the uncertainty of irrational transactions~\cite{QSPHM}: if transactions are completely rational, the stock price should be determined with certainty (and should be equal to the real value of the stock), while the irrationality of transactions will introduce additional fluctuations of the price and thus will lead to a finite small but persistent uncertainty.

The quantum-mechanics way to describe a stock index is naturally analogous to the case of a single stock: we regard the stock index as a particle interacting with a large number of single stocks--a thermal reservoir. The stock index is thus regarded as a system coupled with an environment which carries a tremendous number of harmonic oscillators. The interaction between the index and the stocks is purely linear. The influence of macroscopic external information such as political events and natural disasters is typically considered as a potential well $V(x)$. Because of the huge degrees of freedom of the entire stock market, it is essential for the stock index to be considered as a quantum open system~\cite{OpenQuantumSystems}. Therefore, we can introduce a density operator $\rho =\sum\nolimits_{i}{C_i \left| {{\psi }_{i}} \rangle\langle  {{\psi }_{i}} \right|}$ to describe the stock index. By taking the average of the interaction between the system and environment, we can solve the dynamics of the open system. This is referred to as the \textit{quantum master equation} method~\cite{OpenQuantumSystems}. Treating the Brownian particle as an open system $A$ and the thermal reservoir as an environment $E$ yields the Hamiltonian
\begin{equation}\label{OscillatorHamiltonian}
H={{H}_{A}}+{{H}_{E}}+{{H}_{I}} =\frac{1}{2M}{{P}^{2}}+V(X)+\sum\nolimits_{i}{\left( \frac{1}{2{{m}_{i}}}p_{i}^{2}+\frac{1}{2}{{m}_{i}}\omega _{i}^{2}x_{i}^{2} \right)}-X\sum\nolimits_{i}{{{\kappa }_{i}}{{x}_{i}}},
\end{equation}
where ${\kappa }_{i}$ is the coupling strength between the harmonic oscillator $i$ and the Brownian particle. We use the quantum master equation method on Eq.~(\ref{OscillatorHamiltonian}) and simply derive the Caldeira-Leggett master equation~\cite{CLModel}
\begin{equation}\label{CLeqn}
\frac{d}{dt}{{\rho }_{A}}(t)=-\frac{i}{\hbar}[{{H}_{A}},{{\rho }_{A}}(t)]-\frac{i\gamma }{\hbar}[X,\{P,{{\rho }_{A}}(t)\}]-\frac{2M\gamma kT}{{{\hbar}^{2}}}[X,[X,{{\rho }_{A}}(t)]]
\end{equation}
with its Ohmic-Lorentz spectral density $J(\omega)=\sum\nolimits_{i}{\kappa _{i}^{2}\delta (\omega -{{\omega }_{i}})/2{{m}_{i}}{{\omega }_{i}}}\approx 2{{\pi }^{-1}}M\gamma \omega {{\Omega }_{\text{cut}}^{2}}/\left( {{\Omega }_{\text{cut}}^{2}}+{{\omega }^{2}} \right)$ reflecting the autocorrelation features of the system. The cut-off frequency ${\Omega }_{\text{cut}}$ of the spectral density is, ad hoc though, of no necessity to be considered in Eq.~(\ref{CLeqn}) since the Markovian approximation is already used here. The Markovian approximation is valid only if the characteristic time of the environment is less than the relaxation time, i.e., $\max\{{{\Omega }_{\text{cut}}^{-1}},\hbar /2\pi kT\}\ll {{\gamma }^{-1}}$. We will see later in Section~\ref{NonMarkovianFeaturesOfTheQuantumBrownianMotionModel} that ${\Omega }_{\text{cut}}$ with its specific forms arises in the corresponding non-Markovian master equation of the quantum Brownian motion.

We note that the mapping between a quantum system and the stock market specified here may not be mathematically exclusive. It is of more interest to consider the qBm(m) in the view of statistical physical significance. On the one hand, external information $V(x)$ affects the stock index, by which the information is carried to the environment that consists of large numbers of stocks; on the other hand, the linear response by transactions of stocks to the coming information is fed back to the stock index. The physical dynamics, based on the fluctuation-dissipation theorem, reflects the exchange of energy between the quantum open system and its environment~\cite{QBMFundamentalAspects}.

In summary, the mapping between a quantum system and a stock market is proposed in Tables~\ref{tab1} and~\ref{tab2}.

\begin{center}
\begin{table*}[tbp!]
\caption{ \label{tab1} Mapping between a quantum closed system and a single stock~\cite{PersistentFluctuationsStockMarkets, QSPHM}.}
\begin{tabular}{p{8cm}p{8cm}}

Quantum closed system   &  Single stock    \\
\hline
Coordinate representation $X_i$                     & (logarithmic) Stock price $\ln  {{s}_{i}}$ \\
Momentum representation $P_i$                    & Trend of stock price ${{m}_{i}}d\left( \ln  {{s}_{i}}\right)/dt$\\
Mass ${m}_{i}$                                  & Inertia of stock $i$\\
Energy $E_i$          & Trading volume of stock $i$\\
Wave function (amplitude) $|\psi _{i}(x,t)|^{2}$            & Probability density distribution of stock price\\
Uncertainty relation $[X,P]=i\hbar $              & Uncertainty of irrational transaction\\
\hline
\end{tabular}
\end{table*}

\begin{table*}[btp!]
\caption{ \label{tab2} Mapping between a quantum open system and a stock index.}
\begin{tabular}{p{8cm}p{8cm}}

Quantum open system                                                                      &  Stock index    \\
\hline
Density operator (population) $\rho_{A}(x,x,t)$                                      & Probability density distribution of stock index\\
Potential well $V(x)$                                            & Macroscopic external influence on stock index\\
Thermal reservoir ${{\rho }_{E}}$              & Large numbers of stocks\\
Temperature $kT$                                                                           &Fluctuation strength\\
Dissipation coefficient $\gamma$ of thermal reservoir                   & Dissipation strength\\
Spectral density $J(\omega )$ of thermal reservoir                        & Autocorrelation feature\\
\hline

\end{tabular}
\end{table*}
\end{center}

\section{Moments of the quantum Brownian motion model}
\label{MomentsOfTheQuantumBrownianMotionModel}
\begin{figure}[t]
\begin{centering}
\subfloat[ ]{\begin{centering}
\includegraphics[height=4.4cm]{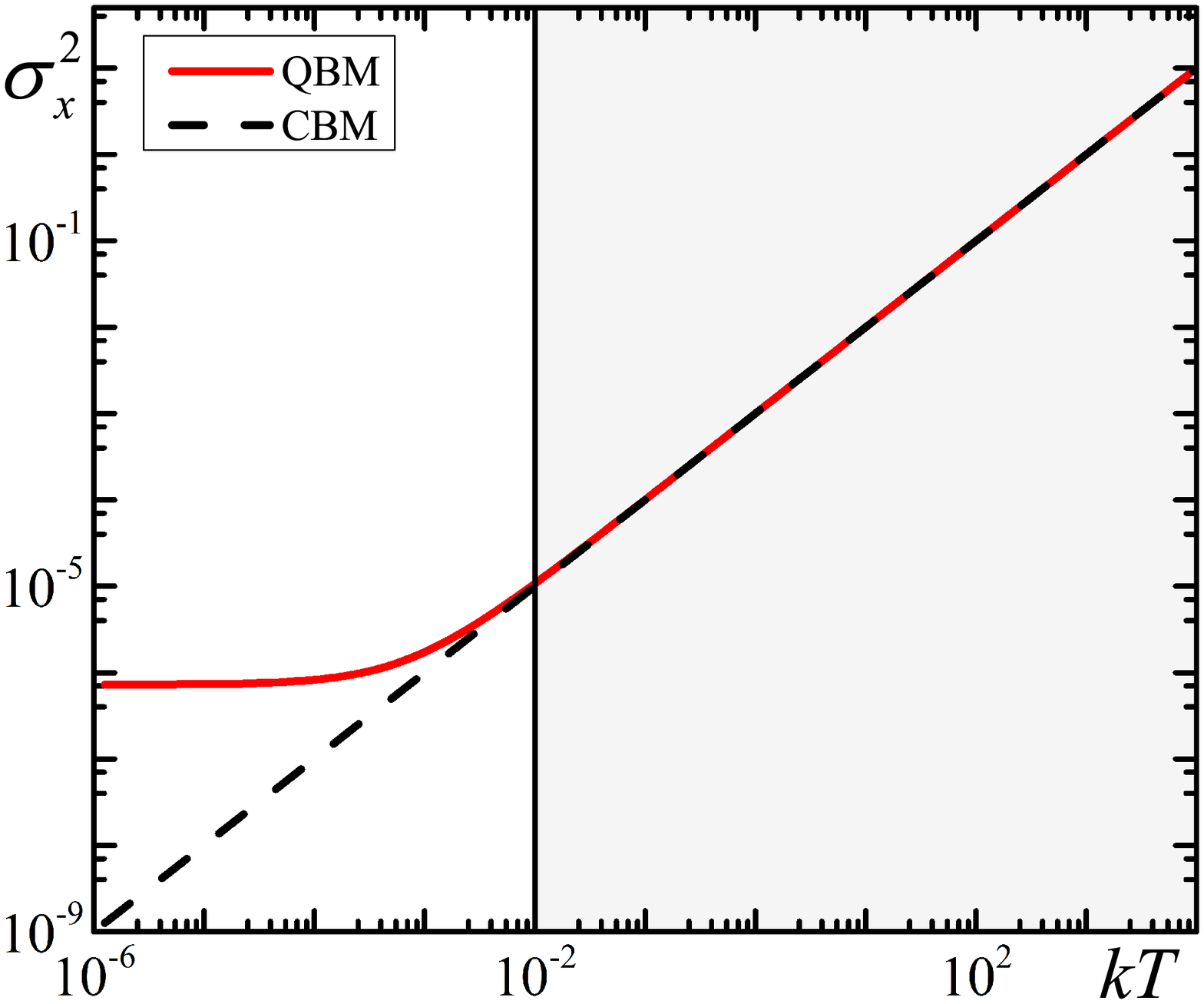}
\par\end{centering}
}\subfloat[ ]{\begin{centering}
\includegraphics[height=4.4cm]{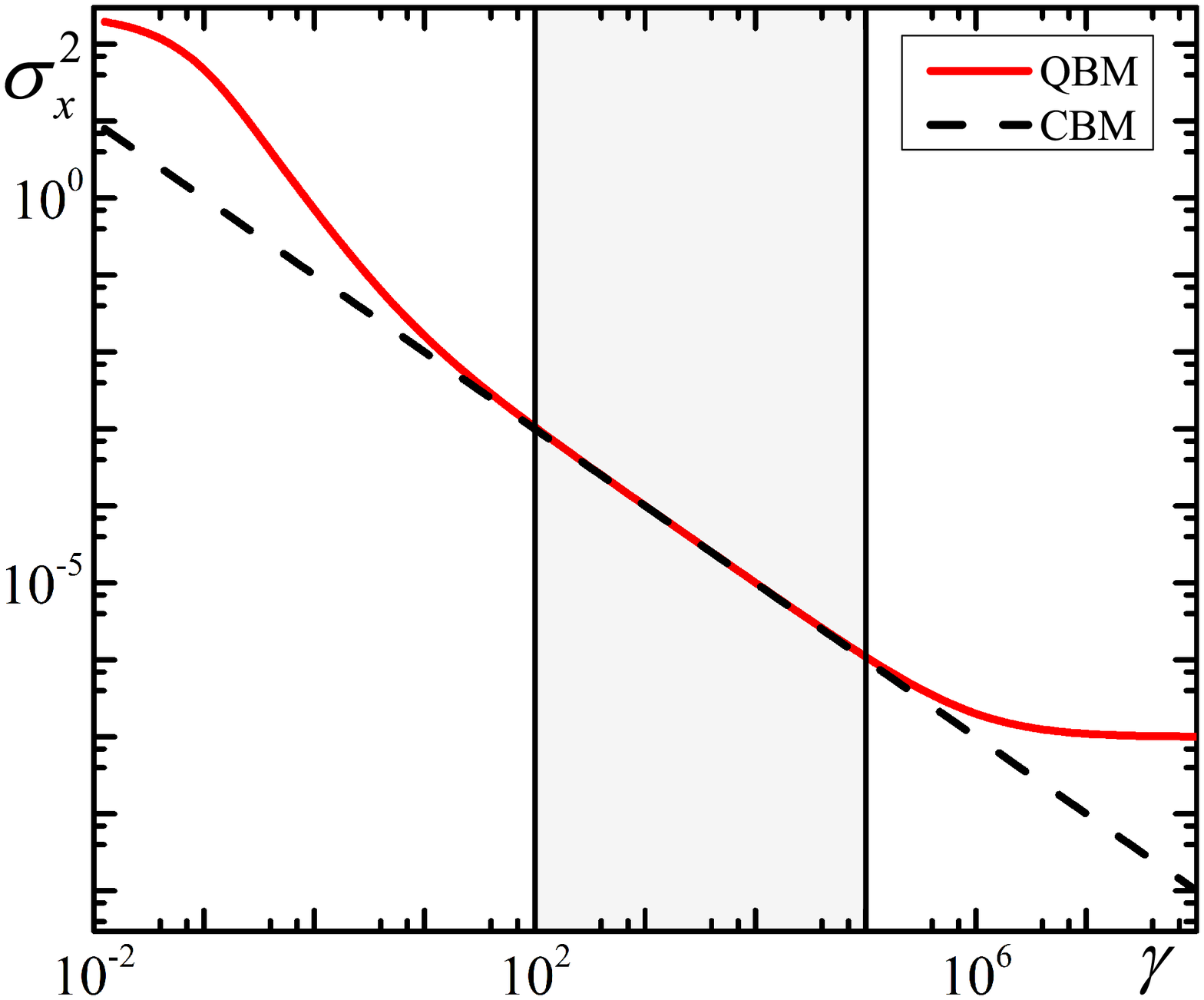}
\par\end{centering}
}\subfloat[ ]{\begin{centering}
\includegraphics[height=4.4cm]{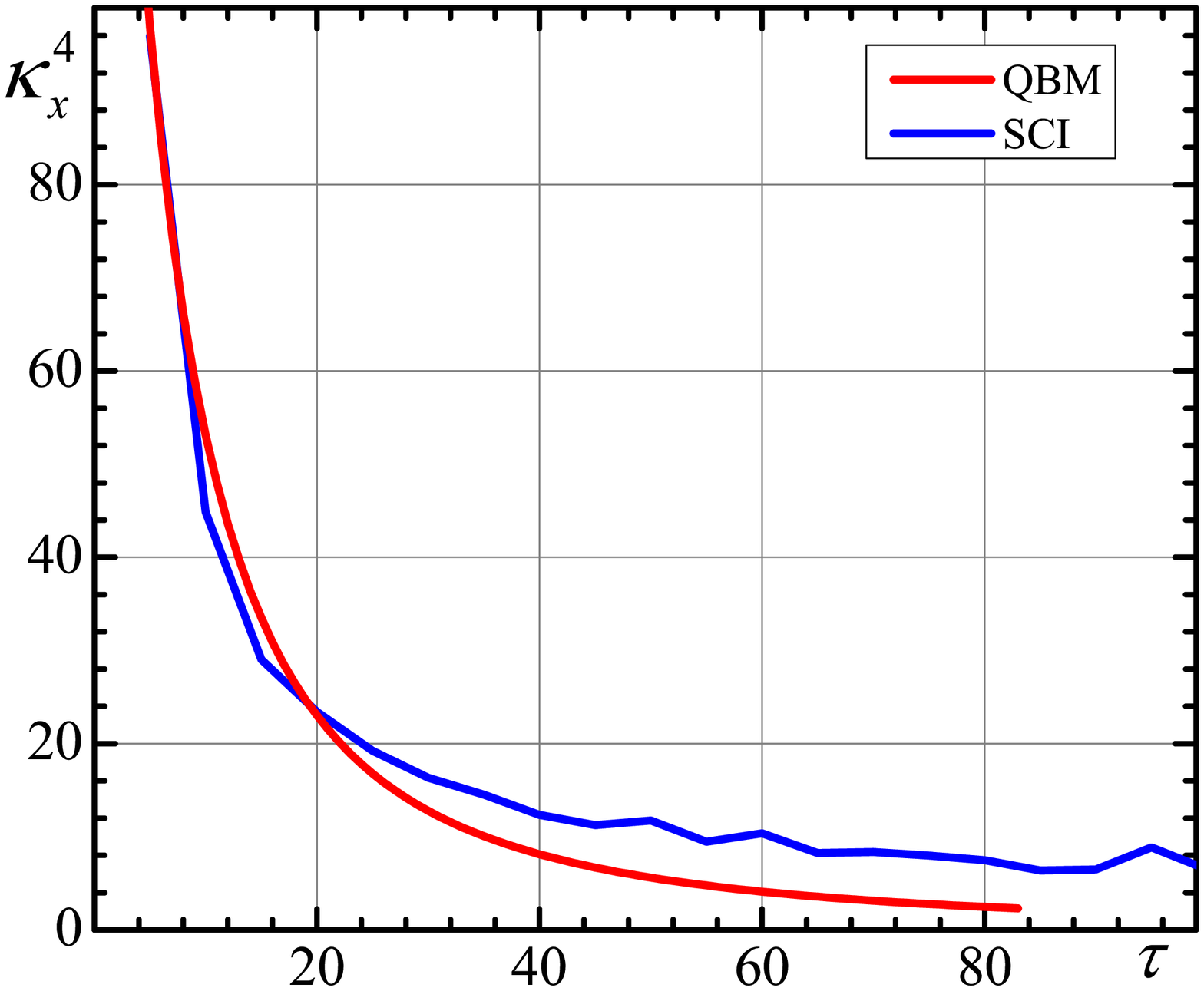}
\par\end{centering}
}
\par\end{centering}
\caption{\label{Fig2}Comparison between the variance of the quantum Brownian motion (red solid line) and the variance of the classical Brownian motion (black dash line) with respect to (a) the temperature $kT$ (when $\gamma=10^{3}$) and (b) the dissipation coefficient $\gamma$ (when $kT=0.1$). $M=10$, $\hbar=0.01$, $t=10$, and $\sigma _{x}^{2}(0)=10^{-7}$ are set for both plots. (c) Comparison between the kurtosis of the quantum Brownian motion (red line) and the kurtosis of the Shanghai Composite Index from 1999 to 2013 (blue line) are shown, with $M=20$ and $kT=\hbar=\gamma=1$. $\left\langle {{X}^{2}} \right\rangle=\left\langle {{P}^{2}} \right\rangle=\hbar/2$, $\left\langle {{XP+PX}} \right\rangle=0$, and $\left\langle {{X}^{4}} \right\rangle=50 \hbar^2$ when $\tau=0$. (For interpretation of the references to color in this figure legend, the reader is referred to the web version of this article.)}
\end{figure}

With the help of Eq.~(\ref{CLeqn}), it is able to calculate the moments of $X$ and $P$ to different orders, e.g., variance, kurtosis, etc. For convenience, we choose the first-order moments $\left\langle X \right\rangle$ and $\left\langle P \right\rangle$ to be zero and all physical parameters to be dimensionless. We have $\sigma _{x}^{2} = \left\langle {X}^{2} \right\rangle$, actually a time-dependent variance,
\begin{equation}\label{sxx}
\sigma _{x}^{2}(t)=\sigma _{x}^{2}(0)+{{\left( \frac{1-{{e}^{-2\gamma t}}}{2M\gamma } \right)}^{2}}\sigma _{p}^{2}(0)+\frac{1-{{e}^{-2\gamma t}}}{2M\gamma }{{\sigma }_{px}}(0)+\frac{kT}{M\gamma }\left[ t+\frac{1}{\gamma }{{e}^{-2\gamma t}}-\frac{1}{4\gamma }\left( {{e}^{-4\gamma t}}+3 \right) \right],
\end{equation}
where $\sigma _{p}^{2} = \left\langle {P}^{2} \right\rangle$ and $\sigma _{px} = \left\langle XP+PX \right\rangle$ ~\cite{OpenQuantumSystems}. For the qBm(m), the volatility of the stock index is influenced by the parameters $kT$ and $\gamma$ in distinct ways (see Eq.~\ref{sxx}); while for the cBm(m), there is only one effective parameter $kT/M\gamma$. The difference of $\sigma _{x}^{2}$ between the qBm(m) and cBm(m) is furthermore compared in Figs.~\ref{Fig2}(a) and~\ref{Fig2}(b).

Under a relatively high temperature $(kT > 10^{-2})$, $\sigma _{x}^{2}(t)$ of the quantum Brownian motion is almost equal to that of the classical Brownian motion; with a low temperature though, the second term in the right side of Eq.~(\ref{sxx}) satisfies $\sigma _{p}^{2}(0)\geq{{\hbar }^{2}}/4\sigma _{x}^{2}(0)$ (due to the uncertainty relation) and cannot be ignored (see Fig.~\ref{Fig2}(a)). Hence, it should be considered that the fluctuations of the stock index have an additional minimum variance caused by irrationality, a dominant factor when $kT \ll \hbar \gamma$. From Fig.~\ref{Fig2}(b), we see that when $ {{10}^{2}}<\gamma <10^5$, the linear dissipation coefficient $\gamma$ leads to $\sigma _{x}^{2}(t) \to (kT/M\gamma)t$. However, if $\gamma > {{10}^{6}}$, the asymptotic behavior is $\sigma _{x}^{2}(t) \to \sigma _{x}^{2}(0)$ instead. A non-zero $\sigma _{x}^{2}(0)$ can be phenomenologically regarded as the noise of the quantum measurement, i.e., an extrinsic variance caused by the imprecision of measurement~\cite{QuantumNoise}, which makes sense if we take the imprecision of calculation of the stock index in actuality into consideration. For an initial state with a minimal uncertainty $\sigma _{p}^{2}(0)={{\hbar }^{2}}/4\sigma _{x}^{2}(0)$, if $(\gamma t\ll 1)$ (the relaxation process cannot be ignored), approximation of Eq.~(\ref{sxx}) becomes $\sigma _{x}^{2}(t)\approx \sigma _{x}^{2}(0)+{{\hbar }^{2}}{{t}^{2}}/{{M}^{2}}\sigma _{x}^{2}(0)+4kT\gamma t^3/3M$~\cite{OpenQuantumSystems}. The third term increases with a speed of $\Theta ({{t}^{3}})$ and denotes the relaxation process; while the second term increases with a speed of $\Theta ({{t}^{2}})$, which, as a unique quantum effect coming from the uncertainty relation, reflects the broadening of a wave packet~\cite{OpenQuantumSystems}.

In reality, even though $\sigma _{x}^{2}$ of the qBm(m) has different behaviors than that of the cBm(m), the good fit of the volatility of SCI with ${\tau}^{1/2}$ (see Fig.~\ref{Fig1}(a)) implies that the difference is negligible, unless in some cases the parameters $kT$ and/or $\gamma$ have extreme values, e.g., when sudden downturn or intense resistance occurs in the stock market. Indeed, a more evident difference can be revealed by higher-order moments: by calculating the kurtosis ${\kappa}_{x}^{4}$ of SCI, we find a non-zero ${\kappa}_{x}^{4}(\tau)$ decreasing exponentially with respect to $\tau$ (see Fig.~\ref{Fig2}(c)). The non-zero ${\kappa}_{x}^{4}(\tau)$ is an evidence of the abnormal deviation from a Gaussian shape for the probability density distribution as shown in Fig.~\ref{Fig1}(b). Giving $\kappa _{x}^{4}(t)=\left\langle {{X}^{4}} \right\rangle /{{\left\langle {{X}^{2}} \right\rangle }^{2}}-3$, it is theoretically implied that the time evolution of the kurtosis of the qBm(m) should depend on initial states distinctly. By several attempts we find that with a non-zero positive $\left\langle {{X}^{4}} \right\rangle$ as the initial condition, ${\kappa}_{x}^{4}(t)$ does evolve decreasingly and exponentially as the actual kurtosis of SCI does (see Fig.~\ref{Fig2}(c)). When $\tau$ is rather large, i.e., when the measurement frequency is low, ${\kappa}_{x}^{4}$ is small enough that the deviation vanishes; in contrast, high-frequency measurements will lead to a noise-type disturbance to ${\kappa}_{x}^{4}$. The dependence of the noise on both initial states and measurement frequency here should not be owing to the imprecision but the back-action noise of measurement~\cite{QuantumNoise} on the stock index. Interestingly, the two types of quantum noise--imprecision and back-action noise--are conjugate as $X$ and $P$ are~\cite{QuantumNoise}. Explanation of the non-Gaussian fat-tail distribution for stock indices is thus in phenomenological concordance to the quantum Brownian motion now if viewed from a perspective of quantum measurement theory.

\section{Non-Markovian features of the quantum Brownian motion model}
\label{NonMarkovianFeaturesOfTheQuantumBrownianMotionModel}
In general, non-Markovian behaviors for both dissipation and fluctuation features of the quantum Brownian motion are reflected by environment-introduced non-Markovian quantum noise~\cite{QuantumNoise}. The frequency respondence of the total quantum noise is asymmetric with respect to $\pm \omega$ and can be divided into a one-sided response function of $+\omega$ (in the role of dissipation) and a "measurable" symmetric function with respect to $\omega$ (in the role of fluctuation), respectively~\cite{QuantumNoise}. From Eq.~(\ref{OscillatorHamiltonian}) we derive~\cite{NoMarkovQBMeqn}
\begin{equation}\label{Nmarkov}
\frac{d}{dt}{{\rho }_{A}}(t)=-\frac{i}{\hbar }[{{H}_{A}},{{\rho }_{A}}(t)]+\mathcal{K}(t){{\rho }_{A}}(t)
\end{equation}
with
\begin{equation}\label{NMarkoveqn}
\mathcal{K}(t){{\rho }_{A}}(t)=-\frac{i\gamma }{\hbar }[X,\{P,{{\rho }_{A}}(t)\}]-\frac{1}{2{{\hbar }^{2}}}\int_{0}^{t}{d\tau {D}_{1}(\tau )[X,[X,{{\rho }_{A}}(\tau )]]}+\frac{1}{2M{{\hbar }^{2}}}\int_{0}^{t}{d\tau \left( \tau {D}_{1}(\tau )[X,[P,{{\rho }_{A}}(\tau )]] \right)}.
\end{equation}
In Eq.~\ref{NMarkoveqn}, if we only consider the "measurable" part and neglect the dissipation part of non-Markovian behaviors of quantum Brownian motion, then $\gamma$ is time-independent. ${D}_{1}(\tau )=2\hbar \int_{0}^{\infty }{d\omega J(\omega )\coth \left( \hbar \omega /2kT \right)\cos }\omega \tau $ is the "measurable" autocorrelation function of fluctuations with respect to $J(\omega)$ of the thermal reservoir~\cite{OpenQuantumSystems}. If ${D}_{1}(\tau )$ is not a Dirac-peak distribution, the memory kernel $\mathcal{K}(t)$ is time-dependent, by which non-Markovian behaviors of quantum Brownian motion are introduced.

\begin{figure}[t]
\begin{centering}
\subfloat[ ]{\begin{centering}
\includegraphics[height=5.5cm]{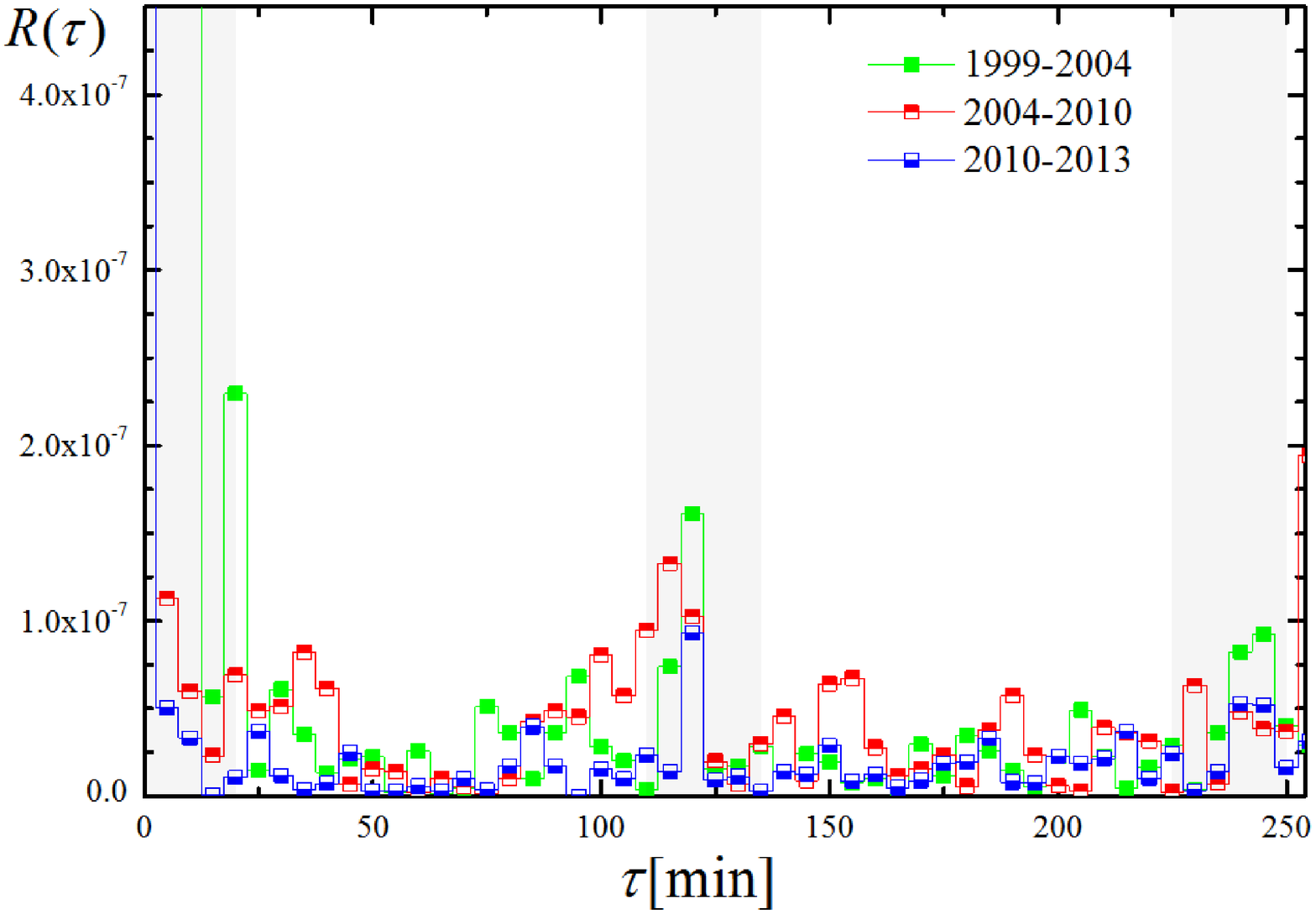}
\par\end{centering}
}
\subfloat[ ]{\begin{centering}
\includegraphics[height=5.5cm]{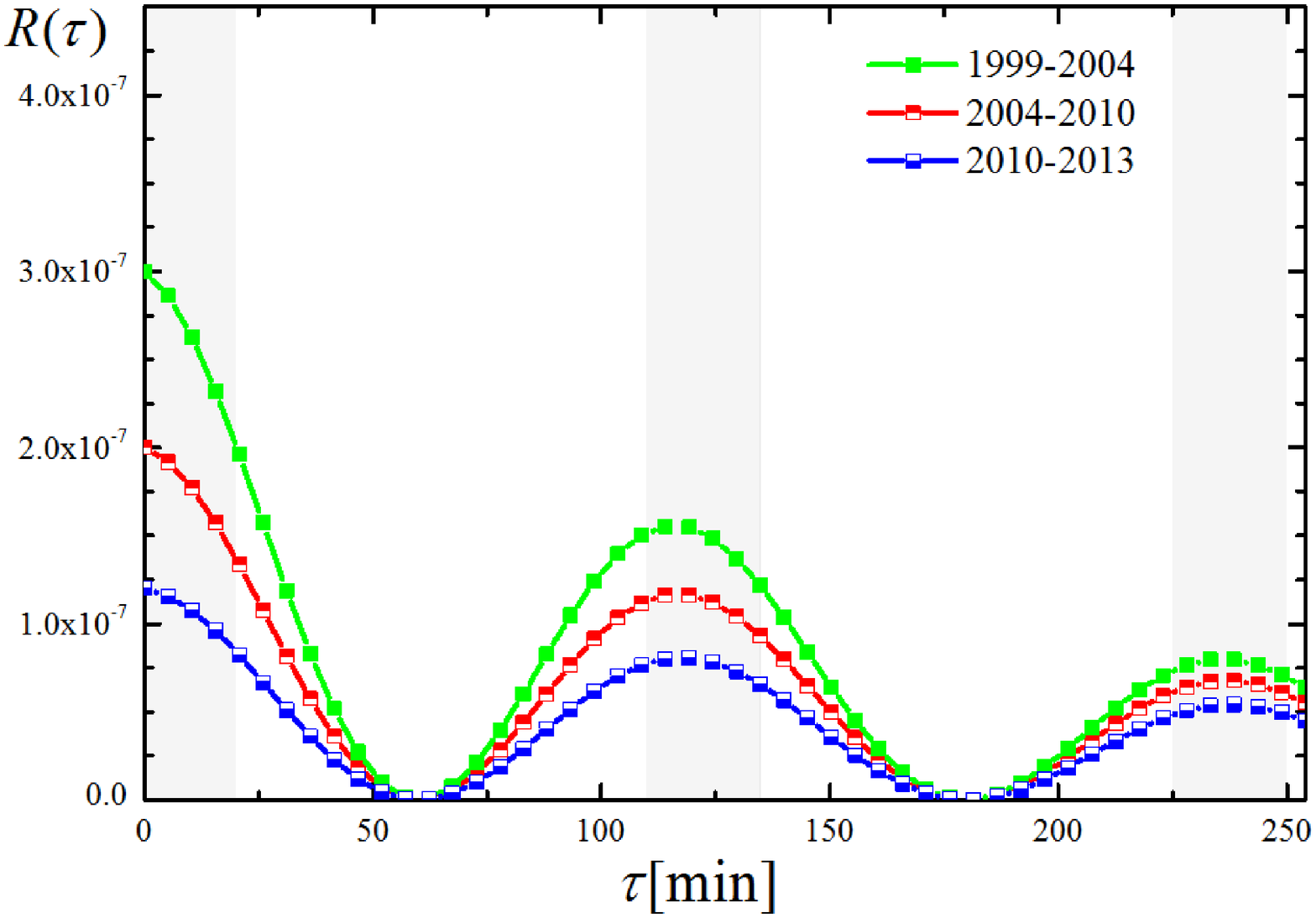}
\par\end{centering}
}
\par\end{centering}
\caption{\label{Fig3} 1999-2004 (green lines), 2004-2010 (red lines), 2010-2013 (blue lines) three periods of the Shanghai Composite Index (SCI), of which (a) the actual autocorrelation of the stock index return is calculated and (b) a simulated non-Markovian autocorrelation $R(\tau)$ is introduced. The fitting parameters are
1999-2004: $\eta=5.56\times {{10}^{-3}}{{\min }^{-1}}, \Omega=8.33\times {{10}^{-3}}\pi {{\min }^{-1}}, \xi=5.48\times {{10}^{-4}}[\ln S]{{\min }^{-1}};$  2004-2010: $\eta=4.55\times {{10}^{-3}}{{\min }^{-1}}, \Omega=8.33\times {{10}^{-3}}\pi {{\min }^{-1}}, \xi=4.47\times {{10}^{-4}}[\ln S]{{\min }^{-1}};$  2010-2013: $\eta=3.33\times {{10}^{-3}}{{\min }^{-1}}, \Omega=8.33\times {{10}^{-3}}\pi {{\min }^{-1}}, \xi=3.46\times {{10}^{-4}}[\ln S]{{\min }^{-1}}.$ (For interpretation of the references to color in this figure legend, the reader is referred to the web version of this article.)}
\end{figure}

In Fig.~\ref{Fig3}(a), non-Markovian behaviors of the stock market are implied by the autocorrelation function $R(\tau)$. Data of three different periods are used for calculating $R(\tau)$. The local maximum points of the three lines all appear at $\tau=120\min$ and $\tau=240\min$ nearby, while the local minimum points mostly appear at $\tau=60\min$ and $\tau=180\min$, indicating a periodic dependence of the autocorrelation of the stock market on $\tau$. It is also shown that $R(\tau)$ decays exponentially in average. In Fig.~\ref{Fig3}(b), a simulated autocorrelation function
\begin{equation}\label{Rtau}
R(\tau )={{\left( \xi {{e}^{-\frac{\eta \tau }{2}}}\cos \Omega \tau  \right)}^{2}}=\frac{1}{2}{{\xi }^{2}}{{e}^{-\eta \tau }}\left( \cos 2\Omega \tau +1 \right)
\end{equation}
is introduced to fit the non-Markovian autocorrelation functions of the stock market. $\xi$ denotes the intensity of non-Markovian effect, $\eta$ the decay rate, and $\Omega$ an introduced frequency for the periodicity of the entire stock market. ${D}_{1}(\tau )$ consists of two independent terms now,
\begin{equation}\label{Htau}
{D}_{1}(\tau )=8M\gamma kT\delta (\tau )+8{{M}^{2}}{{\gamma }^{2}}R(\tau ).
\end{equation}
Substituting Eq.~(\ref{Htau}) into Eq.~(\ref{NMarkoveqn}) and taking the time-convolutionless perturbation approximation~\cite{TCLMasterEquation, NMarkov2} to the second order (${{\rho }_{A}}(\tau )\to {{\rho }_{A}}(t)$), we eventually derive
\begin{equation}
\label{UltimateEqn}
\mathcal{K}(t){{\rho }_{A}}(t)=-\frac{i\gamma }{\hbar }[X,\{P,{{\rho }_{A}}(t)\}]-\Delta (t)[X,[X,{{\rho }_{A}}(t)]]+\Lambda (t)[X,[P,{{\rho }_{A}}(t)]],
\end{equation}
where
\begin{equation}
\Delta (t)=\frac{2M\gamma kT}{{{\hbar }^{2}}}+2{{\left( \frac{M\gamma \xi }{\hbar } \right)}^{2}}\left\{ \frac{1}{\eta }\left( 1-{{e}^{-\eta t}} \right)+\frac{\eta }{{{\eta }^{2}}+4{{\Omega }^{2}}}\left[ 1-{{e}^{-\eta t}}\left( \cos 2\Omega t-\frac{2\Omega }{\eta }\sin 2\Omega t \right) \right] \right\},
\end{equation}
and
\begin{eqnarray}
\Lambda (t)&=&\frac{2M{{\gamma }^{2}}{{\xi }^{2}}}{{{\hbar }^{2}}{{\eta }^{2}}}\left[ 1-\left( 1+\eta t \right){{e}^{-\eta t}} \right]\nonumber\\
&&+\frac{2M{{\gamma }^{2}}{{\xi }^{2}}}{{{\hbar }^{2}}\left( {{\eta }^{2}}+4{{\Omega }^{2}} \right)} \left\{\frac{{{\eta }^{2}}-4{{\Omega }^{2}}}{{{\eta }^{2}}+4{{\Omega }^{2}}}-{{e}^{-\eta t}}\left[ \left( \frac{{{\eta }^{2}}-4{{\Omega }^{2}}}{{{\eta }^{2}}+4{{\Omega }^{2}}}+\eta t \right)\cos 2\Omega t +\left( \frac{4\eta \Omega }{{{\eta }^{2}}+4{{\Omega }^{2}}}+2\Omega t \right)\sin 2\Omega t \right] \right\}.
\end{eqnarray}
Furthermore, the asymptotic solution of $J\left(\omega\right)$ is
\begin{equation}
\label{Jw}
J\left(\omega\right) \simeq \frac{2M \gamma}{\pi} \omega+ \frac{M^2 \gamma^2 \xi^2 \eta }{\pi kT} \left[ \frac{2}{\eta^2+\omega^2}+\frac{1}{\eta^2+\left(\omega-2\Omega\right)^2}+\frac{1}{\eta^2+\left(\omega+2\Omega\right)^2} \right] \omega
\end{equation}
when $\omega \to 0$. The second term of Eq.~(\ref{Jw}) consists of three different Ohmic-Lorentz cut-off functions and contributes to the non-Markovian behaviors.

Equation~(\ref{Nmarkov}) with a $\mathcal{K}(t)$ following Eq.~(\ref{UltimateEqn}) is an appropriate non-Markovian dynamic equation for the Shanghai stock market. It is worth noting that the observed autocorrelation of the stock market must be fundamental in the qBm(m). The autocorrelation comes directly from the spectral density $J(\omega)=\sum\nolimits_{i}{\kappa _{i}^{2}\delta (\omega -{{\omega }_{i}})/2{{m}_{i}}{{\omega }_{i}}}$, which introduces the weight of contribution from a specific single stock oscillator ${\omega}_{i}$ to the stock index. With an Ohmic spectral density $J(\omega_i )= 2 M\gamma \omega_i / \pi$, the weight is exactly proportional to the energy quanta $\hbar\omega_i /2$ of stock $i$, which results ${D}_{1}(\tau )=8M\gamma kT \delta (\tau)$, a sign for a Markovian stock index. Instead, with an Ohmic-Lorentz spectral density like Eq.~(\ref{Jw}), the stock index is non-Markovian apparently. It is implied that one can understand the correspondence between stocks and their index through non-Markovian behaviors of the index itself.

Equation~(\ref{Rtau}) is simply mono-frequency. Because of linearity of interactions we can attach higher-order terms with frequencies of $2\Omega, 3\Omega ...$ for a more sophisticated calculation. Additionally, one should notice that the autocorrelation function should yield $R(\tau)e^{-i\phi}$ actually. Nevertheless, one can only concern on the amplitude $R(\tau)$ by assuming the phase information ${\phi}$ a completely time-uncorrelated quantity.

\section{Discussion and Conclusion}
\label{Conclusion}
Exact non-Markovian solution of Eq.~(\ref{OscillatorHamiltonian}) indicates that the master equation contains not only second-order terms like Eq.~(\ref{NMarkoveqn}) but also higher-order products of $X$ and $P$~\cite{QBMExactMasterEquation}. By introducing the Wigner function $ {{{W}}_{\rho }}(x,p,t)=\int_{-\infty }^{+\infty }{ds}{{\rho }_{A}}\left( x-s/2,x+s/2,t \right)\exp \left( ips/\hbar  \right)$ in phase space, a Kolmogorov equation analogous to Eq.~(\ref{KKeqn}) can be derived~\cite{Kolmogorov},
\begin{equation}\label{Keqn}
\frac{\partial }{\partial t}{{W}_{\rho }}(x,p,t)=-\frac{p}{M}\frac{\partial }{\partial x}{{W}_{\rho }}+2\gamma \frac{\partial }{\partial p}\left( p{{W}_{\rho }} \right)+{{\hbar }^{2}}\Delta (t)\frac{{{\partial }^{2}}}{\partial {{p}^{2}}}{{W}_{\rho }}-{{\hbar }^{2}}\Lambda (t)\frac{{{\partial }^{2}}}{\partial x\partial p}{{W}_{\rho }}+\mathcal{G}({{W}_{\rho }}),
\end{equation}
where
\begin{equation}\label{HigherOrderCorrection}
\mathcal{G}({{W}_{\rho }})=\sum\nolimits_{k=3}^{\infty }{\sum\nolimits_{r=0}^{k}{\frac{{{\left( -1 \right)}^{k}}}{r!\left( k-r \right)!}\frac{{{\partial }^{k}}\left[ {{A}^{\left( k-r,r \right)}}{{W}_{\rho }} \right]}{\partial {{x}^{k-r}}\partial {{p}^{r}}}}}
\end{equation}
with time-dependent coefficients ${{A}^{\left( k-r,r \right)}\left( x,p,t \right)}=\langle {{\left[ x(t+dt)-x(t) \right]}^{k-r}} \rangle \langle {{\left[ p(t+dt)-p(t) \right]}^{r}} \rangle / dt$ ($dt\to 0$). Equation~(\ref{HigherOrderCorrection}) is responsible for correction to the long-term solution of Eq.~(\ref{Keqn}), which is not guaranteed to be Gaussian if the $\mathcal{G}({{W}_{\rho }})$ term exists. Now it seems invalid since the stochastic dynamics of any stock index should satisfy a Gaussian distribution in the long-term limit. Nevertheless, it is proved by the Pawula theorem~\cite{ImpossibleNonGaussian} that if any ${{A}^{\left( k-r,r \right)}}\left( x,p,t \right)$ is zero, we must have all ${{A}^{\left( k-r,r \right)}}\equiv0$ for $r\ge 3$. $\mathcal{G}({{W}_{\rho }})$ therefore equals zero in most conditions, guaranteeing that the long-term solution of ${W}_{\rho }$ is Gaussian.

Another possible confusion about the qBm(m) would be concerned with the redundance of completely using the mathematical structures of quantum mechanics. It is already shown that the uncertainty relation (or, more generally, the Lie algebra structure) of noncommutative operators is an appropriate frame to quantify transaction irrationality~\cite{PersistentFluctuationsStockMarkets, QSPHM}. However, we have no idea of whether the non-diagonal elements (coherences) of a density operator, as additional degrees of freedom, have any practical meaning or not. Instead of considering coherences as only redundant auxiliary variables, we rather believe that coherences are some/a kind of hidden variables describing the correlation between individual transactions, which is in concordance with the expectation of synchronized behaviors in behavioral economics~\cite{SimpleMarketModelsFail}.

To summarize, for the purpose of overcoming certain limitations of the efficient market hypothesis and classical Brownian motion model, a quantum Brownian motion model (qBm(m)) is introduced for the stock market (Shanghai Stock Exchange of China is typically studied in this paper). Analogous mappings from a quantum closed/open system to the stock market are proposed with the help of quantum open system theory (see Tables~\ref{tab1} and~\ref{tab2}). With the Caldeira-Leggett master equation used, moments of the qBm(m) to different orders are calculated for comparison with the classical model. The non-zero kurtosis of the stock index (Shanghai Composite Index) that suggests a non-Gaussian fat-tail distribution is fitted by the qBm(m). Explanation details involve the concepts of imprecision and back-action noise in the quantum measurement theory. To describe non-Markovian behaviors of the stock market, a non-Markovian master equation is obtained. The spectral density of autocorrelation is directly related to the weight of the contribution from single stocks to the stock index. With the qBm(m) for the stock index constructed, the entire stock market can be modeled integrally into the quantum-mechanics framework~\cite{QSPHM} now. Our future research aim will concentrate on both theoretical and empirical studies, especially the application of the model to quantitative finance as well as other topics in econophysics.

\section*{Acknowledgements}
We thank Xiumei Yu and Lingze Zhang for their valuable advice and help. We also acknowledge the financial support of the 2012 Challenge Cup Technology Engineering Office of Peking University.

\bibliographystyle{model1-num-names_AuthorFullName}
\bibliography{QBM}







\end{document}